\documentclass[prd, twocolumn, nofootinbib]{revtex4}

\usepackage{epsfig} 
\usepackage{amsmath}

\newcommand{\beq}{\begin{equation}}
\newcommand{\eeq}{\end{equation}}
\newcommand{\beqa}{\begin{eqnarray}}
\newcommand{\eeqa}{\end{eqnarray}}
\newcommand{\om}{\Omega_m}
\newcommand{\omw}{\Omega_w}

\newcommand{\lam}{\Lambda} 
\newcommand{\wtot}{w_{\rm tot}}

\newcommand{\vp}{V_\phi}

\newcommand{\op}{\Omega_\phi}
\newcommand{\gs}{\gtrsim} 
\newcommand{\ls}{\mathrel{\raise0.27ex\hbox{$<$}\kern-0.70em \lower0.71ex\hbox{{
$\scriptstyle \sim$}}}}

\begin{document} 

\title{Field Flows of Dark Energy} 
\author{Robert N.\ Cahn, Roland de Putter, Eric V.\ Linder} 
\affiliation{Lawrence Berkeley National Laboratory \& University 
of California, Berkeley, CA 94720, USA} 
\date{\today}

\begin{abstract} 
Scalar field dark energy evolving from a long radiation- or matter-dominated 
epoch has characteristic dynamics.  While 
slow-roll approximations are invalid, a well defined field 
expansion captures the key aspects of the dark energy evolution during much 
of the matter-dominated epoch.  Since this behavior is determined, it
is not faithfully represented if priors for dynamical 
quantities are chosen at random.  We demonstrate these features for 
 both thawing and freezing fields, and for some modified gravity models, 
and unify several special cases in the literature. 
\end{abstract} 


\maketitle

\section{Introduction \label{sec:intro}}

The acceleration of the cosmic expansion can be explained by a
multitude of models suggested in the literature but few or none
derived from first principles. We look, therefore, for common
characteristics among classes of models and for physically motivated
generic behaviors.  One example of effective 
characterization is the designation of the thawing and freezing
patterns of dark energy evolution \cite{caldlin}.  These involve
scalar fields leaving or approaching the behavior of a
cosmological constant in the dark energy equation of state $w$ and
its time variation $w'=dw/d\ln a$.  Here we seek further unifying
features in the evolution of the dark energy field. 

For inflation a slow-roll approach, neglecting higher time derivatives
in the field evolution, can be used.  This is not valid for dark
energy, even at early times when $w$ may be near $-1$ and the dark 
energy is a small contribution to the 
total energy density, and we must develop a different formalism.
Several suggestions for specific approximations exist in the
literature and we will see that these can be unified into a single
approach.  This can also be extended in part to the freezing class of 
fields, traditionally difficult to characterize generically.

One key conclusion is that one must take into account that our universe 
is old: the scalar field has been evolving for many Hubble times in a 
background that was initially radiation-dominated then matter-dominated. 
This defines particular initial conditions and determines the dynamical 
behavior.  Employing random, Monte Carlo initial conditions may lead to 
underrepresentation of thawing or freezing behavior, due to 
neglecting the physics of a long evolution. 

In \S\ref{sec:dyn} we show why slow roll conditions are invalid for
dark energy and discuss new methods for evaluating the dynamics.  We 
provide a specific example of a complete solution in \S\ref{sec:difq}. 
Identifying a particular characteristic combination of parameters in 
\S\ref{sec:flow}, we show how the physics constrains the dark energy 
evolution within the classes of dark energy.  We investigate extending 
this relation to some modified gravity scenarios in \S\ref{sec:Halpha}.

\section{Field Dynamics \label{sec:dyn}} 

Cosmic acceleration is given by the condition $\ddot a>0$ on the 
scale factor $a$, where dots represent time derivatives.  For a 
 state of near exponential expansion, the Hubble parameter 
$H=\dot a/a$ follows 
\beq 
|\dot H/H^2|\ll1, 
\eeq 
so the Hubble parameter is nearly constant.  If the scalar field $\phi$ 
provides the dominant contribution to the expansion, as it does for 
inflation, then the potential must be nearly constant, leading to an 
equivalent condition $|V_\phi/V|\ll1$ where $V_\phi=dV/d\phi$ and we 
work in units with $8\pi G=1$.  This is 
often referred to as the (first) slow-roll parameter.  Another implication 
is that the field acceleration $\ddot\phi\ll 3H\dot\phi$ and so can be 
neglected in the equation of motion, or Klein-Gordon equation.  

\begin{figure}[!htb]
\begin{center} 
\psfig{file=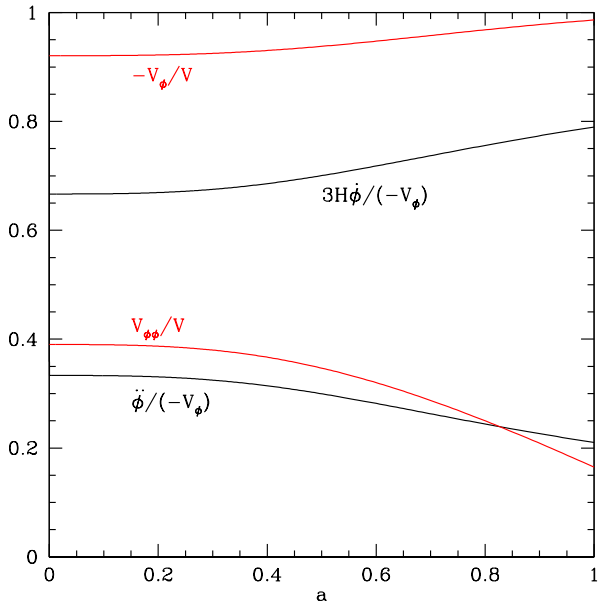,width=3.4in} 
\caption{Even a model evolving only from $w(z\gg1)=-1$ to $w_0=-0.9$ 
(here a PNGB 
model with $f=1$, see \S\ref{sec:thawevo}) cannot be described 
by a slow-roll formalism.  Generically, none of the ratios of terms in 
the scalar field equation of motion, or the potential derivatives, is 
small enough to permit neglect of a term or to allow use of the 
slow-roll approximation. 
}
\label{fig:terms} 
\end{center} 
\end{figure}

However, these slow-roll conditions on $V$ or $\ddot\phi$ rely on scalar 
field domination, and this is not valid for the dark energy field 
during its evolution, even today (since the matter density $\om$ is not 
negligible).  
In \cite{paths}, ratios of terms in the Klein-Gordon equation were 
defined as 
\beqa 
X&\equiv&\frac{\ddot\phi}{H\dot\phi}, \\ 
Y&\equiv&\frac{\ddot\phi}{\vp}=\frac{-X}{X+3}, 
\eeqa 
and shown today (when $\om\approx0.28$) to be far from a slow-roll regime. 
In Fig.~\ref{fig:terms} we show explicitly that even for a thawing 
model that has equation of state today $w_0=-0.9$, near the de Sitter 
$w=-1$ value, 
no term in the equation of motion is negligible 
compared to other terms, and also that $|\vp/V|\ll1$ does not hold, 
at any point in the evolution.

\subsection{General} 

With all terms retained, the Klein-Gordon equation for 
the scalar field $\phi$ is given by 
\beq 
\ddot\phi+3H\dot\phi=-\vp\,. \label{eq:kg}
\eeq 
The ratios $X$, $Y$ of the terms are of interest for several reasons: 
1) they indicate whether the potential driving term or Hubble friction 
term is more influential (as used in \cite{caldlin} to motivate thawing 
and freezing behaviors), 2) the ratios can be rephrased in terms of the 
tracks in the equation-of-state phase space $w'$-$w$, and 3) under certain 
conditions, such as tracking (constant $w$ determined by the 
background expansion \cite{zws,swz}), $X$ and $Y$ will be constant.  

To begin, we obtain an implicit solution by multiplying Eq.~(\ref{eq:kg}) 
through by an integrating factor $\exp\{3\int_\tau^t dt'\,H(t')\}$, 
where the lower limit is arbitrary.  Integrating from some early time 
$t_1$ to a later time $t_2$ yields 
\beqa 
\dot\phi(t_2)&-&e^{3\int_{t_2}^{t_1} dt'\,H(t')}\,\dot\phi(t_1) \nonumber \\ 
&=&-\int_{t_1}^{t_2} dt\,e^{3\int_{t_2}^{t} dt'\,H(t')}\,\vp(t).  
\eeqa 
Note that 
\beq 
e^{3\int_{t_2}^{t_1} dt'\,H(t')}=(a_1/a_2)^3\,, \label{eq:ht}
\eeq 
so for $t_1$ early enough relative to $t_2$ this factor is very small 
and the term involving $\dot\phi(t_1)$ is irrelevant.  That is, the 
initial speed does not matter due to the high Hubble drag in the early 
universe.  This leaves (taking the scale factor $a_1\to0$) 
\beqa 
\dot\phi(t)&=&-\int_{-\infty}^{t} dt'\,e^{3\int_{t}^{t'} dt''\,H(t'')} 
\,\vp(t') \nonumber \\ 
&=&-\int_0^a \frac{da'}{a'H(a')}\left(\frac{a'}{a}\right)^3\,\vp(a'). 
\label{eq:phidot} 
\eeqa 

Note that the ratio of the friction term and the driving term in the 
Klein-Gordon equation can be evaluated as 
\beq 
\frac{3H\dot\phi}{\vp}=-3\int_0^a \frac{da'}{a'}\left(\frac{a'}{a}\right)^3\,
\frac{H(a)}{H(a')}\frac{\vp(a')}{\vp(a)}. 
\eeq 
It is also convenient to consider the ratio of the field acceleration 
to the driving term, 
\beqa 
Y\equiv\frac{\ddot\phi}{\vp}&=&\frac{-\vp-3H\dot\phi}{\vp} \nonumber \\ 
&=&3\int_0^a \frac{da'}{a'}\left(\frac{a'}{a}\right)^3\,
\left[\frac{H(a)}{H(a')} \frac{\vp(a')}{\vp(a)}-1\right]. \label{eq:yint}
\eeqa 
This is a formal solution because both $H$ and $V_\phi$ 
depend on $\phi$ itself. 

\subsection{Asymptotic} 

Some instructive cases can be evaluated directly. Consider a 
model where $\vp$ is nearly constant in an epoch dominated by a 
component with equation of state $w_b$, so that $H\sim a^{-3(1+w_b)/2}$.  
This is generic for the thawing class.  Evaluating Eq.~(\ref{eq:yint}) 
gives 
\beq 
Y_\infty=-\frac{1+w_b}{3+w_b}, \label{eq:yinf} 
\eeq 
or $Y_\infty=-1/3$ for the matter dominated era, $Y_\infty=-2/5$ for 
the radiation dominated era.  

\subsection{Thawing Evolution \label{sec:thawevo}} 

The evolution of the thawing field can be determined iteratively from 
Eq.(\ref{eq:phidot}), taking into account that initially the matter 
contribution dominates in $H$ and that the field moves relatively little 
from its initial value.  Taylor expanding the potential around that 
initial field value (which without loss of generality can be set to zero), 
\beqa
V(\phi)&=&V_i+V_{\phi i}\phi+\frac 12V_{\phi\phi i}\phi^2+\ldots \\ 
H^2&=&\frac 13[\rho_m(a)+\rho_\phi(a)]=
\frac 13\left[\rho_m(a)+V_i+\ldots\right] \,, 
\eeqa 
where the subscript $i$ denotes the value at $\phi=\phi_i=0$, 
$\rho_m$ is the matter density, and $\rho_\phi$ is the dark energy 
density\footnote{One 
can avoid specifying $H$ explicitly by rewriting the Klein-Gordon equation as 
\beq
\phi'' + \frac{3}{2} \left(1+\Omega_{\phi} - (1+w)\Omega_{\phi}\right) \phi' 
+ 3 (1 - \Omega_{\phi}) \frac{V_{\phi}(\phi)}{\rho_m(a)}=0, \nonumber 
\eeq
where a prime denotes $d/d\ln a$.  Since $\op$, $1+w$, and 
$V_\phi/\rho_m(a)$ are all small at early times, this form is 
convenient for identifying the order of each term.}. 

At lowest order we find
\beqa
\dot\phi(a)&=&-\frac 29\frac{V_{\phi i}}{H_0\Omega_m^{1/2}}\,a^{3/2} \\
\phi(a)&=&-\frac 2{27}\frac{V_{\phi i}}{H_0^2\Omega_m}a^3 
=-\frac 29\frac{V_{\phi i}}{\rho_m(a)} \,.
\eeqa
In second order 
\beqa
\dot\phi(a)&=&-\frac 29\frac{V_{\phi i}a^{3/2}}{H_0\Omega_m^{1/2}} 
\left[1-\frac 35\left(\frac{V_i}{2\rho_m(a)}+\frac{2 V_{\phi\phi i}}{9\rho_m(a)}\right)\right] \label{eq:phidotexp} \\ 
\phi(a)&=& -\frac{2}{9} \frac{V_{\phi i}}{\rho_m(a)} \left[1 - \frac{2}{5} \left(1 + \frac{V_{\phi \phi i}}{6 V_i}\right)
\frac{V_i}{\rho_m(a)}\right]. \label{eq:phiexp} 
\eeqa 
Thus the criterion for the validity of the first order solution for 
$\phi$ is mainly that $V_i/\rho_m(a)\ll1$, i.e.\ dark energy does not 
dominate, as expected.  There are no slow roll $V_{\phi}/V\ll1$ or 
$V_{\phi\phi}/V\ll1$ conditions or other stringent condition on the 
potential derivatives.  Indeed one could argue that having $V$ and 
its derivatives be of the same order in Planck units is natural, as 
in the technically natural PNGB model \cite{Friemanetal95} with 
$V=V_0\,[1+\cos(\phi/f)]$ where $f$ is a symmetry energy scale of 
order unity.  If $V_{\phi\phi}/V\gg1$, we see from 
Eqs.~(\ref{eq:phidotexp})--(\ref{eq:phiexp}) we might have to 
reevaluate the field expansion; see for example the next subsection.  

From these expressions we can calculate $Y$ directly, substituting in 
the first order corrections to $H$, $\phi$, and $V$: 
\beq 
Y(a)=-\frac 13+\left(\frac 2{15}+\frac 8{135}\frac{V_{\phi\phi i}}{V_i}\right)\frac{V_i}{\rho_m(a)} \,. 
\eeq 
The parameter $Y$ starts off constant, 
at $Y_\infty$, only changing 
as the scalar field rolls sufficiently far or its energy density 
starts to dominate the Hubble expansion.  This behavior is evident in 
Fig.~\ref{fig:terms}.  Also note that $Y$ is not 
particularly small, and hence no one term dominates in the Klein-Gordon 
equation and ignoring $\ddot\phi$ is invalid.  

It is similarly straightforward to determine the leading correction 
to the dark energy equation of state $w$ by substituting the first 
order expressions into 
\beq
w=\frac{\frac 12\dot\phi^2-V}{\frac 12\dot\phi^2+V}
\eeq
to find
\beq
1+w=\frac4{27}\frac{V_{\phi i}^2}{V_i\rho_m(a)} \,. \label{eq:eos}
\eeq
The time variation of the equation of state is 
\beq 
w'=3(1+w), \label{eq:wwp3} 
\eeq 
the result obtained by \cite{caldlin}.  That is, the evolution of the 
field in a matter-dominated universe fixes the asymptotic behavior 
of the dark energy for such thawing fields.

\subsection{Non-analytic Potentials} 

As alluded to above, when the derivatives of the potential become large 
at a point, the Taylor expansion approach can break down.  Consider a class 
of potentials with a singularity at some $\phi_\star$, e.g. 
\beq
V=V_\star-V_n(\phi-\phi_\star)^n\,, 
\eeq
with $n$ non-integral and positive (we discuss negative $n$ below) 
and $\phi>\phi_\star$.  This 
represents an inverted (concave) potential with the field rolling away 
from a maximum at $\phi=\phi_\star$ (eventually to negative infinity but 
that will not concern us regarding early time behavior).  

We can find a thawing solution by trying 
\beq 
\phi=\phi_\star+Aa^\nu 
\eeq 
in the Klein-Gordon equation during the matter-dominated era (easily 
generalized to other background evolution), or 
equivalently into Eq.~(\ref{eq:phidot}).  The result is 
\beqa
\nu&=&\frac 3{2-n}\\
A&=&\left[\frac{V_n}{H_0^2\Omega_m}\frac{2n(2-n)^2}{9(4-n)}\right]^{1/(2-n)}\,. 
\eeqa 
The ratio of the field acceleration to potential slope terms, the equation 
of state, and derivative of the equation of state become 
\beqa  
Y&=&\frac{-n}{4-n} \\ 
1+w&=&\frac{H_0^2\Omega_m}{V_\star}A^2\nu^2 a^{3n/(2-n)} \\ 
w'&=&\frac{3n}{2-n}(1+w) 
\eeqa 
as the field starts to roll.  

Note that for $n>2$, in this Ansatz the field starts 
with a large kinetic energy, or equivalently $w$ is positive and large, 
so we restrict to $0\le n<2$ (for $n=2$ the Ansatz fails).  As $n\to2$, 
the dark energy shoots away from $w=-1$, acting more like sublimation 
than thawing.  For the two integer values of $n$ within this range, 
$n=0$, 1, the potential has no singularities and 
these results agree with the Taylor expansion of the previous subsection. 
As well, if the field starts frozen away from the singularity then Taylor 
expanding the field works and the $w'=3(1+w)$ trajectory is the 
early-time solution. 

If instead we consider negative $n$, we can ignore the $V_\star$ term at 
early times and (making the potential convex) we end up with a tracking 
field \cite{ratrapeebles}.  The equations for $\nu$, $A$, and $Y$ above 
still hold but now 
\beq 
w_{n<0}=-\frac 2{2-n}\,. 
\eeq 
and thus $w'=0$.

\subsection{Unifying Relations \label{sec:unify}} 

Not only the dynamical trajectories but the relations between the dark 
energy density and the equation of state have characteristic behavior 
for each class of models.  Combining Eq.~(\ref{eq:eos}) with the first 
order expression for the dark energy density, 
\beq
\label{eq:Om V}
\Omega_{\phi} = \frac{V}{\rho_m(a)}\,, 
\eeq
shows that 
\beq
\label{eq:Om w}
\Omega_{\phi} = \frac{27}{4} \frac{1+w}{\lambda^2}\,, 
\eeq 
where $\lambda=-V_{\phi}/V$, 
here considered to lowest order in an expansion in $\phi$.  
(Note that generally $\lambda^2\equiv 2\epsilon$, where $\epsilon$ is the 
conventional first slow-roll parameter.)  We return to this relation 
between the parameters in \S\ref{sec:flow}; now we simply explore some 
implications of the existence of such a relation. 

Note that to first order the relation between $\Omega_{\phi}$ and 
$w$ does not depend on higher derivatives of $V$ than the first 
derivative, and the relation between $w'$ and $w$ does not depend 
on $V$ or its derivatives at all to this order.  This is part of the 
unifying power of such an analysis, that {\it any\/} scalar field 
dominated by the background Hubble friction must behave in a simple, 
determined manner. 

We can now compare our result to other approaches in the literature 
that assumed specific potentials or approximations.  
The simplest case is the limit $V =$ constant or $\lambda \to 0$.  
This is of course just the cosmological constant and the equation of 
state never leaves $w=-1$.  (Skating models \cite{liddleskate,linderskate} 
have a large initial motion $\dot\phi_i$ but this quickly redshifts 
away, as $a^{-3}$, and the field comes to rest at $w=-1$.)  
Next is the linear potential \cite{linde86,Kalloshetal03}, 
where $\vp=$ constant, discussed in the next section.  In general, though, 
potentials will have higher order derivatives that are not zero or depend 
nontrivially on the first derivative.  

Thawing models have been studied with approximations, such as 
taking $\lambda=$ constant (turning the exponential potential's tracking 
behavior \cite{wett88,Copeetal98} into thawing by starting it from a 
frozen state) but approximating 
$\Omega_\phi$ or $w$ \cite{ScherrerSen08}.  Indeed \cite{ScherrerSen08} 
noted a version of the relation (\ref{eq:Om w}) then holds asymptotically.  
Another parameter investigated in a first order expansion about a constant 
value is $\kappa\equiv-\lambda/(1 + X/3)$ \cite{CrittMajPiaz07}. 
Explicitly incorporating Eq.~(\ref{eq:wwp3}), 
\cite{lindyn} expanded in the energy density $\op$ about the asymptotic 
solution to form an ``algebraic thawing'' model, which is actually 
valid to second order.  Interestingly, current data show the algebraic 
thawing model is statistically a better fit than $\lam$CDM \cite{rubin}. 

All of these cases follow the unifying first order solution (\ref{eq:Om w}) 
but diverge at higher 
order.  Each one basically chooses different ways to truncate or close 
the hierarchy of higher order equations.  To understand how quickly 
deviation from the unified solution, or of the validity of the field 
expansion approach, occurs, we note it requires particular combinations of 
$\Omega_{\phi}$ and $1+w$ to be much smaller than one.  So we do not 
expect these analytic field solutions to be valid up to the 
present.  Nevertheless, in many cases they are good approximations until 
surprisingly recent times; e.g.\ for the model of Fig.~\ref{fig:terms} 
the relationship (\ref{eq:Om w}) holds to 3\% (8\%) until $z=2$ (1).  
We discuss this further in \S\ref{sec:flow}.

\section{Dark Energy Density and the Linear Potential\label{sec:difq}} 

In addition to understanding the dark energy equation of state, 
we are often interested in the observables directly, such as the 
Hubble parameter or the distance-redshift relation.  These involve the 
dark energy density, given by the sum of the potential $V(\phi)$ and 
kinetic energy $\dot\phi^2/2$ discussed in the previous section. 

The dark energy density at some epoch relative to its current value is 
given by the ratio 
\beq 
\xi\equiv\frac{\rho_\phi(a)}{\rho_{\phi,0}}= 
\frac{\frac 12\dot\phi^2+V}{\frac 12\dot\phi_0^2+V_0} \,, 
\eeq 
where the subscript $0$ indicates current values.  The current value for 
the equation of state is 
\beq 
w_0=\frac{\frac 12\dot\phi^2_0-V_0}{\frac 12\dot\phi_0^2+V_0} \,. 
\eeq 
Differentiating the dark energy density with respect to time and using the 
Klein-Gordon equation, we find 
\beq 
\frac{d\xi}{dt}=-\frac{3H\dot\phi^2}{V_0}\,\frac{1-w_0}{2} 
\eeq 
or 
\beq 
\frac{d\xi}{da}=-\frac{3\dot\phi^2}{V_0}\,\frac{1-w_0}{2a} \,. 
\eeq 
To remove the explicit appearance of $\dot\phi$ we use 
\beq 
1+Y=-\frac{3H\dot\phi}{V_\phi}\,,
\eeq 
and to eliminate $V_0$ we employ 
\beq 
\frac{\dot\phi_0^2}{V_0}=\frac{V_{\phi,0}^2\,(1+Y_0)^2}{9H_0^2V_0} 
=\frac{2(1+w_0)}{1-w_0} \,, 
\eeq 
with the result 
\beq
\frac{d\xi}{da}=-3\frac{1+w_0}{a}\,\frac{H_0^2}{H^2}\, 
\frac{V_{\phi}^2}{V_{\phi,0}^2} 
\left(\frac{1+Y(a)}{1+Y_0}\right)^2 \,. \label{eq:de} 
\eeq 

Note that the Hubble parameter is given by 
\beq 
H^2/H_0^2=\Omega_{m0}\,a^{-3}+(1-\Omega_{m0})\,\xi(a) 
\eeq 
so the density $\xi(a)$ is an implicit function only of the scale factor $a$ 
-- except for the dependence on the potential slope $V_\phi(\phi(a))$. 
This dependence occurs for $Y(a)$ as well through Eq.~(\ref{eq:yint}). 

In the special case of a potential linear in $\phi$, the quantity 
$V_\phi$ is constant.  This was 
treated by \cite{linde86} as one of the first dark energy models and more 
recently as a textbook example by \cite{wbgbook}.  Then the two 
equations (\ref{eq:de}) and 
(\ref{eq:yint}) involve only the independent variable $a$ and the dependent 
variable $\xi$.  
(This will be a good approximation for any potential with 
a slope varying with $a$ sufficiently slowly.)  
Despite being coupled integro-differential equations, they are actually 
simple to solve numerically.  One starts with an initial 
approximation for the Hubble parameter with $\xi$ set to unity for all 
values of $a$, that is initially $H(a)$ is appropriate to a cosmological 
constant.  In this case, two parameters fix the solution: 
$\Omega_\lam=1-\Omega_{m0}$ and $w_0$.  

\begin{figure}[!htb] 
\begin{center} 
\psfig{file=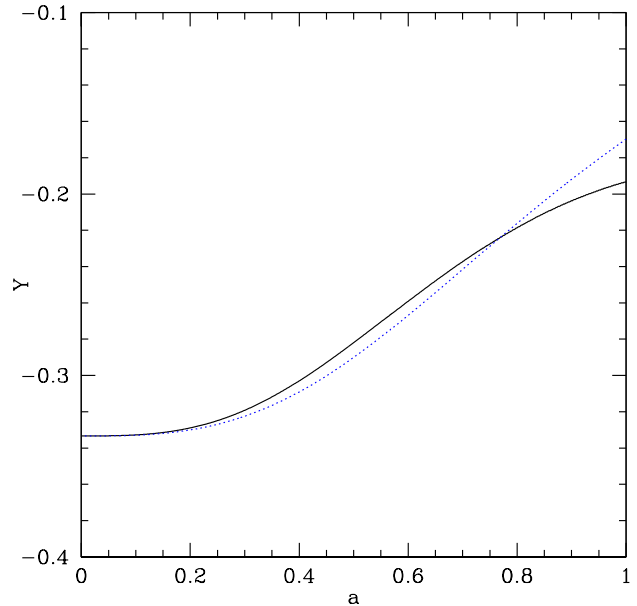,width=3.4in} 
\caption{The acceleration response function $Y(a)$ for any thawing field 
has common characteristics.  Here we plot the case of a linear potential 
with $w_0=-0.777$, $\Omega_{\phi,0}=0.76$ (compare to the PNGB case shown 
in Fig.~\ref{fig:terms}).  
The solid curve is the exact solution, while 
the dotted curve shows the analytical approximation for 
$w_0=-1$, $\Omega_{\phi,0}=0.76$, which is used as the starting 
point for the iterative solution. 
}
\label{fig:gfun} 
\end{center} 
\end{figure}

\begin{figure}[!htb]
\begin{center} 
\psfig{file=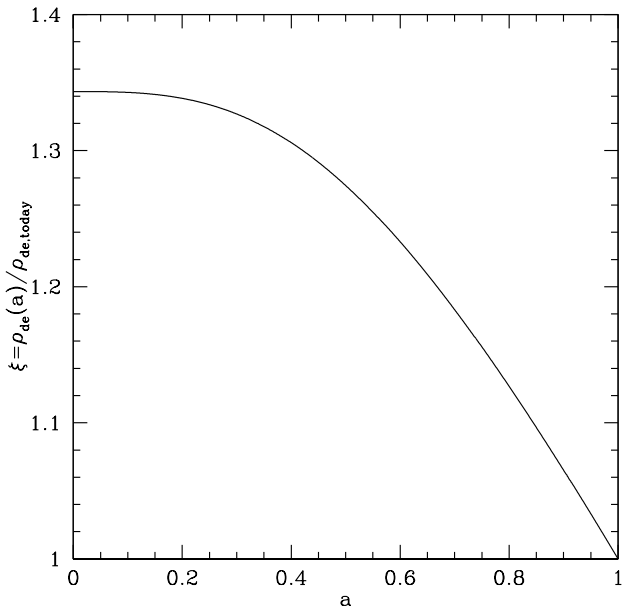,width=3.4in} 
\caption{The dark energy density relative to the present, $\xi(a)$, 
changes little for thawing fields.  The curve shows the solution for 
the linear potential with $w_0=-0.777$, $\Omega_{\phi,0}=0.76$; note the 
density used in the first iteration (dotted curve in Fig.~\ref{fig:gfun}) 
is just the cosmological constant behavior, unity for all scale factors. 
By contrast, for a typical tracking field $\xi(a\ll1)\gg1$, e.g.\ 
$\xi(a=10^{-3})\approx 10^4$ for $V\sim\phi^{-2}$. 
}
\label{fig:xi} 
\end{center} 
\end{figure}

The value of $Y(a)$ is not known until the solution is found 
iteratively, but for the first iteration the cosmological constant 
Ansatz 
\beq 
H(a)/H_0=\sqrt{(1-\Omega_\lam)a^{-3}+\Omega_\lam}\equiv h(a) 
\eeq 
gives 
\begin{eqnarray}
1+Y(a)&=&\frac{1-\Omega_\lam}{\Omega_\lam}\,\frac{h(a)}{a^3} 
\Bigg[\frac{h(a)}{h(a)^2-\Omega_\lam} \nonumber\\ 
&& \qquad+\frac 1{2\Omega_\lam^{1/2}}\ln\frac{h(a)-\Omega_\lam^{1/2}}{h(a)
+\Omega_\lam^{1/2}}\Bigg] \,. 
\end{eqnarray} 
The dark energy density calculated with the first order quantities is 
quite accurate, even when $w_0$ is not very close to $-1$, and 
convergence upon iteration is extremely rapid.  
By taking a further derivative of $\xi$, one can also derive the 
relation for the time variation of the dark energy equation of state 
followed by the linear 
potential, 
$w'_0\equiv dw/d\ln a(a=1)\approx -1.3\,(1+w_0)$.  
Results found iteratively for the linear potential are shown for $Y(a)$ 
and the density $\xi(a)=\rho_\phi(a)/\rho_{\phi,0}$ in 
Figs.~\ref{fig:gfun} and \ref{fig:xi}.

\section{Flow Parameter \label{sec:flow}} 

As we saw in \S\ref{sec:unify}, 
evolution away from the frozen state involves a synchronized deviation in 
both $\Omega_\phi$ and $1+w$ away from zero in a particular proportional 
or scaling manner that persists to quite late times.  
This imposes a constraint that 
phenomenological models must obey, and parameter priors or Monte Carlo 
scans must take 
into account.  Other forms of dark energy, such as modified 
gravity, might also follow similar scaling relations (see, e.g., 
\cite{lincahn}) and we investigate this in the next section.  
Such constraints are important as we lack first principles theories for 
dark energy, as scalar fields or modified gravity. 

\begin{figure}[!htb]
\begin{center} 
\psfig{file=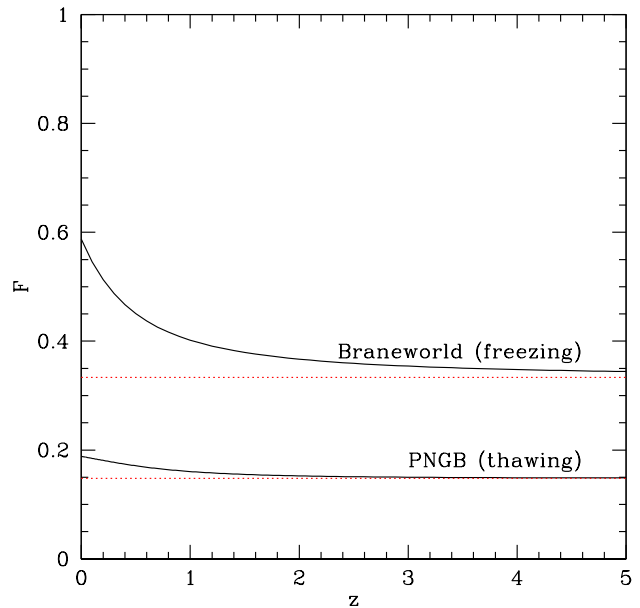,width=3.4in} 
\caption{A particular combination of dark energy parameters, 
$F=(1+w)/[\op(dV/d\phi)^2]$, has fixed high redshift behavior for each 
class of models, $4/27$ for thawing and $1/3$ for freezing (red, dotted 
lines).  Also, it is nearly conserved during the long evolution until 
dark energy begins to dominate. 
}
\label{fig:flowz} 
\end{center} 
\end{figure}

Motivated by the relation in Eq.~(\ref{eq:Om w}), we define 
a new quantity we call the flow parameter, 
\beq
F\equiv \frac{1+w}{\op\lambda^2}\,, \label{eq:fdef} 
\eeq 
where $\lambda=-V_\phi/V$, $\op$, and $w$ are all functions of time. 
This is related to the friction vs.\ potential slope term ratio $1+Y$ by 
\beq 
F=\frac{1}{12}(1-w)^2(1+Y)^2, \label{eq:fy} 
\eeq 
and to the phase space evolution of dark energy as 
\beq 
w'=-3(1-w^2)[1-(3F)^{-1/2}]\,. \label{eq:wwpf} 
\eeq 
These relations follow generally from the Klein-Gordon equation and 
from the definition of $Y$, for all $a$ and without any requirement 
for matter domination.  
We note that $F$ holds close to its high-redshift, asymptotic value for 
a substantial part of the evolution, even when $\op$ is not much less 
than one.  Figure~\ref{fig:flowz} shows the evolution of $F$ for 
thawing and freezing (in fact modified gravity) examples.

As seen in Eq.~(\ref{eq:Om w}), for any thawing model evolving in a matter 
dominated universe the field flow begins with 
\beq 
F=4/27 \qquad {\rm(thawer)}\,. \label{eq:fthaw} 
\eeq 
This is equivalent to the condition $w'=3(1+w)$ 
at early times, but is preserved for more of the evolution.  
The other major class of scalar field models is tracking models.  
These have constant equation of state 
at early times in a background dominated universe, and from  
Eq.~(\ref{eq:wwpf}) this implies 
\beq 
F=1/3 \qquad{\rm(tracker)}. \label{eq:ftrack} 
\eeq 
This can be viewed as $Y(a\ll1)=(1+w_\infty)/(1-w_\infty)$, where 
$w_\infty$ is the high redshift dark energy equation of state. 
The non-analytic thawing solutions with $0<n<1$ interpolate between 
the regular thawing case (cf.\ $n=1$) and tracking models ($n<0$): 
\beq
F=\frac 43\left(\frac{2-n}{4-n}\right)^2\,. 
\eeq
 
Again, there is a substantial range of validity for $F$ being constant 
even as $\op$ grows to an appreciable fraction. 

This relation imposes a particular high redshift behavior on a whole class 
of phenomenological models.  Effectively, these flow behaviors define 
physical priors that must be included in any Monte Carlo simulation 
of scalar field evolution.  As shown, they hold until quite recent 
epochs, $z\gs 1-2$, not just for $a\ll1$.  Assuming instead a random 
dynamical state, rather 
than that determined during the many matter-dominated e-folds of 
expansion, is equivalent to a fine tuning of the scalar field initial 
conditions to avoid the natural evolutionary path.  See 
\cite{CrittMajPiaz07} for an interesting analysis of the effect of the 
difference in priors between those uniform in $w'$-$w$ and those uniform 
in a version of the field parameters.

\section{Modified Gravity and Expansion \label{sec:Halpha}} 

The accelerating expansion of the universe could indicate a deviation 
from general relativity rather than the presence of a new scalar field.  
It is of great interest to explore beyond scalar fields and see whether 
the evolution of expansion and matter density growth parameters 
probes modified gravity.  For example, \cite{lincahn,caldmelch} motivated 
a deviation from the general relativistic growth behavior scaling as 
$\Omega_w(a)$, or $a^3$ in many cases, 
while \cite{amenks} assumed a variation as $a$, and \cite{bertschinger} left 
it as a fit parameter $a^s$.  As seen in \cite{bertschinger}, the 
observational consequences for the deviation index $s$ are significant. 
This is too large and complex a subject to address here, but we examine 
how some modified gravity theories respond to the parameters 
treated above for scalar field explanations of dark energy. 

For gravity extended beyond Einstein relativity, the parameters $\op$, 
$w$, $V$, etc. are effective quantities but we can still define them in 
terms of modifications of the Friedmann equation from the general 
relativistic, matter-dominated case.

\subsection{DGP Braneworld and $H^\alpha$ \label{sec:bwhalpha}} 

We first consider the Dvali-Turner \cite{DvaliTurn03} modification 
(also see \cite{freese03}) 
\beq
H^2 = 8 \pi G \rho_m(a)/3 + (1 - \Omega_{m, 0})\, H_0^2\, (H/H_0)^{\alpha},  
\eeq 
where $\Omega_{m,0}=\Omega_m(a=1)$.  
This was motivated by consideration of extradimensional theories, with 
the index $\alpha$ depending on boundary conditions between our four 
dimensional universe and a higher dimensional bulk volume.  The 
DGP braneworld cosmology 
\cite{dgp,ddg} is the special case $\alpha=1$.  Following \cite{paths}, 
this acts as an effective freezing scalar field with 
\beqa 
w&=& -\left(1 + \frac{\alpha}{2-\alpha} \Omega_m\right)^{-1} \\ 
w'&=& 3 w (1 + w) \left[1 - \frac{2}{\alpha}(1 + w)\right]\,. \label{eq:wpbw}
\eeqa 
This provides the opportunity to follow the flow parameter for a freezing 
(indeed tracking) field starting far from $w=-1$, rather than for a thawing 
field.  From Eq.~(\ref{eq:wwpf}), we find the effective flow parameter 
\beqa
F(a)&=&\frac{1}{3}\,\left[\frac{1-w}{1-(2/\alpha)\,w(1+w)}\right]^2 
\nonumber \\ 
&=& \frac{1}{3}\, \frac{\left(2 \frac{2-\alpha}{\alpha}+\Omega_m\right)^2 
\left(\frac{2-\alpha}{\alpha}+\Omega_m\right)^2}{\left[\Omega_m^2+2(1+\alpha^{-1})\frac{2-\alpha}{\alpha}\Omega_m+\left(\frac{2-\alpha}{\alpha}\right)^2\right]^2}\,. 
\eeqa 

Considering the early time limit of $F$, we expect in the matter-dominated 
universe that $\Omega_m \to 1$, so $F \to 1/3$.  That is, the flow 
parameter indeed agrees with the tracking value of Eq.~(\ref{eq:ftrack}).  
Asymptotically, the effective equation of state is constant at 
$w = -1 + \alpha/2$ and one can in fact show that the effective potential 
has the form of an inverse power law potential $V\sim\phi^{-n}$ with index 
$n=2\alpha/(2-\alpha)$.  

\begin{figure}[!htb]
\begin{center} 
\psfig{file=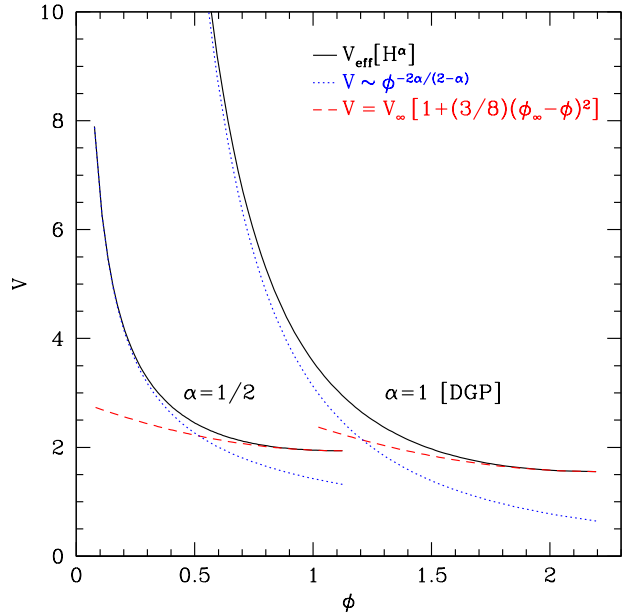,width=3.4in} 
\caption{The effective potential of the modified gravity case involving 
$H^\alpha$ is plotted in units of $H_0^2$ for $\alpha=1/2, 1$.  At 
high redshift (small $\phi$) it possesses the flow parameter of a 
tracking scalar field and acts like an inverse power law potential, 
while in the far future it freezes to a cosmological constant state, 
approaching it as a quadratic potential.  
}
\label{fig:veffbw} 
\end{center} 
\end{figure}

In the future, the $H^{\alpha}$ term will come to dominate (assuming 
$\alpha < 2$) so that $\Omega_m \to 0$. Its effective equation of state
then approaches $w = -1$. The flow parameter $F\to 4/3$, which 
corresponds to $w'=3w(1+w)$.  
This flow parameter is the maximum value for 
freezing scalar fields and the minimum value for barotropic fields 
(see \cite{scherrer06} for the barotropic case).  In this limit, the 
$H^\alpha$ effective potential approaches a quadratic form, 
$V\approx V_\infty\,[1+(3/8)(\phi_\infty-\phi)^2]$, where the 
field has an asymptotic maximum value $\phi_\infty$ and $V_\infty=3H_0^2
\,[1-\om(z=0)]^{2/(2-\alpha)}$.  Note that this class of models thus 
effectively incorporates a cosmological constant.  Figure~\ref{fig:veffbw} 
shows the full solution for the effective potential.

\subsection{$f(R)$ Gravity \label{sec:fR}} 

Another class of theories where dark energy is an effective quantity 
arising from extending gravity is $f(R)$ theories, where the action 
involves a function of the Ricci scalar, here considered in addition 
to the usual linear term (so $f=0$ corresponds to general relativity). 
\cite{songhs} gives the modified Friedmann expansion equation and an 
expression for the effective dark energy density $\rho_{\rm de}$.  
We consider the deviation from the early, high curvature, 
matter-dominated era and write the equation in terms of the effective 
total equation of state of universe $\wtot$, 
\beqa 
f''&+&\left(\frac{1}{2}+\frac{3}{2}\wtot+W\right)f' \nonumber\\ 
&-&\frac{3}{2}(1-2\wtot-3\wtot^2+\wtot')\,f \nonumber \\ 
&=&24\pi G\,\rho_{\rm de}\, (1-2\wtot-3\wtot^2+\wtot')\,,  
\eeqa 
where $W=(3-3\wtot-15\wtot^2-9\wtot^3+5\wtot'+9\wtot\wtot'-\wtot'')/ 
(1-2\wtot-3\wtot^2+\wtot')$. 

Now we consider the evolution of the departure from general relativistic 
matter domination with total effective equation of state $\wtot=0$. 
Since the total equation of state $\wtot=w\Omega_w(a)$, where the 
dimensionless effective dark energy density $\Omega_w(a)= 
(8\pi G/3)\rho_{\rm de}/H^2$, then we have $\Omega_w(a)$, $\wtot$, and its 
derivatives all of the same order.  The dark energy density $\omw(a)\ll1$ 
at the epoch considered so we can expand in this small quantity. 
The zeroth order solution, with $\wtot=0=\omw(a)$, gives $f\sim a^p$ 
with $p=(-7+\sqrt{73})/4$, as found by \cite{songhs}.  This homogeneous 
solution acts as an initial condition that quickly becomes unimportant. 
The first order 
departure from matter domination and general relativity has 
\beq 
f_1(a)=A\,H^2(a)\,\omw(a), 
\eeq 
where $A$ is of order unity, and shows that gravity deviates from 
general relativity at the same rate as the effective dark energy density 
evolves (cf.\ \cite{lincahn}).  The proportionality quantity $A$ does 
not have to be constant, in general, just of order unity (thus any time 
variation is on Hubble scales or longer). 

If we restrict $A$ to be constant, then we can solve the equation in 
terms of effective dark energy equation of state and phrase this as 
the flow parameter 
\beq 
F(a)=\frac{3(1-w^2)^2}{[3A^{-1}-2-(5/2)w]^2}\,. 
\eeq 
We can find a tracking solution with $w'=0$, yielding $A=-3/[1-(5/2)w-3w^2]$, 
so a given choice of constant $A$ corresponds to a given constant $w$. 
Since the dark energy density evolves as $\rho_{\rm de}\sim a^{-3(1+w)}$ 
and the Ricci scalar is dominated by matter $\rho_m\sim a^{-3}$, in this 
limit the gravitational modification looks like $f(R)\sim R^{1+w}$. 
However, considerable freedom exists to choose other solutions, e.g.\ 
with $A$ varying.  

Note that the physics governing the true scalar field evolution can be 
very different from that operating for modified gravity, so there is no 
expectation that the same relations should hold. 
Flows unlike the well-determined quintessence 
behavior may provide hints of modified gravity.

\section{Conclusions \label{sec:concl}} 

Our universe is old, having expanded by a factor of perhaps $10^{28}$ 
since the last period of acceleration during the inflationary epoch. 
This strongly affects the evolution of a scalar field that 
may give rise to the current epoch of acceleration and determines some 
key properties of the dark energy.  Although conventional slow roll 
approximations are invalid, we show that analysis in terms of an 
integral relation between the Hubble friction and potential driving terms 
and a well characterized field expansion can give insights into the 
evolutionary behavior. 

For the case of thawing fields, the field expansion provides a clear 
initial track in the $w$-$w'$ phase space, a unification of a number 
of interesting special cases, and a rapid convergence in the evolution 
of the dark energy density.  For tracking fields, the ratio between 
the Klein-Gordon terms reaches a constant value.  In both cases, the 
evolution of the deviation of $\op$ from zero and of the tilt $1+w$ 
scale in a manner constrained by the long matter (and radiation) 
dominated era.   Phrased in terms of a flow parameter combining the 
scalings, this ratio is nearly conserved until quite recent times, 
$z\approx1-2$, when the dark energy finally begins to take over.  

This physical behavior means that dark energy dynamics is not random, 
or equally probable, e.g.\ in the sense of a uniform prior over $w$-$w'$, 
but is focused -- ``flows'' -- in specific ways. 
We have also tested this for two modified gravity theories and found some 
similar behavior but also some deviations that could offer clues to 
the nature of the acceleration.   While one can always arrange initial 
conditions such that the dark energy comes out of the matter-dominated 
era with arbitrary behavior, this involves fine tuning.  The oldness of 
our universe does provide a natural path for dark energy dynamics.

\acknowledgments 

We thank Steven Weinberg for motivating part of this work and Wayne Hu 
for useful discussions.  EL and RdP are grateful to the Michigan Center 
for Theoretical Physics for hospitality.  
RNC thanks Institut National de Physique Nucl\'eaire et de Physique des 
Particules for support and the Laboratoire de Physique Nucl\'eaire et 
Hautes Energies for its hospitality.  
This work has been supported in part by the Director, Office of Science, 
Office of High Energy Physics, of the U.S.\ Department of Energy under 
Contract No.\ DE-AC02-05CH11231.


\begin{thebibliography}{99}

\bibitem{caldlin} 
R.R. Caldwell \& E.V. Linder 2005, Phys. Rev. Lett. 95, 141301

\bibitem{paths} 
E.V. Linder 2006, Phys. Rev. D 73, 063010

\bibitem{zws} 
I. Zlatev, L. Wang, P.J. Steinhardt, P.J. 1999, Phys. Rev. Lett. 82, 896 

\bibitem{swz} 
P.J. Steinhardt, L. Wang, I. Zlatev, I. 1999, Phys. Rev. D 59, 123504

\bibitem{Friemanetal95} 
J.A. Frieman, C.T. Hill, A. Stebbins, I. Waga 1995, Phys. Rev. Lett. 75, 2077 

\bibitem{ratrapeebles} 
B. Ratra \& P.J.E. Peebles 1988, Phys. Rev. D 37, 3406

\bibitem{liddleskate} 
M. Sahl{\'e}n, A.R. Liddle, D. Parkinson 2005, Phys. Rev. D 72, 083511 

\bibitem{linderskate} 
E.V. Linder 2005, Astropart. Phys. 24, 391 

\bibitem{linde86} 
A. Linde 1987, in {\it Three Hundred Years of Gravitation\/}, ed.\ S W 
Hawking \& W Israel (Cambridge: Cambridge U.\ Press), p.\ 604

\bibitem{Kalloshetal03} 
R. Kallosh, J. Kratochvil, A. Linde, E.V. Linder, M. Shmakova 2003, JCAP 
0310, 015 

\bibitem{wett88} 
C. Wetterich 1988, Nucl. Phys. B 302, 668 

\bibitem{Copeetal98} 
E.J. Copeland, A.R. Liddle, D. Wands 1998, Phys. Rev. D 57, 4686

\bibitem{ScherrerSen08} 
R.J. Scherrer \& A.A. Sen 2008, Phys. Rev. D 77, 083515 

\bibitem{CrittMajPiaz07} 
R. Crittenden, E. Majerotto, F. Piazza 2007, Phys.\ Rev.\ Lett.\ 98, 251301 

\bibitem{lindyn} 
E.V. Linder 2008, Gen.\ Rel.\ Grav.\ 40, 329 [arXiv:0704.2064] 

\bibitem{rubin} 
D. Rubin et al., arXiv:0807.1108

\bibitem{wbgbook} 
S. Weinberg 2008, Cosmology (Oxford U. Press) 

\bibitem{lincahn} 
E.V. Linder \& R.N. Cahn 2007, Astropart. Phys. 28, 481 

\bibitem{caldmelch} 
R. Caldwell, A. Cooray, A. Melchiorri 2007, Phys. Rev. D 76, 023507

\bibitem{amenks} 
L. Amendola, M. Kunz, D. Sapone 2007, JCAP 0804, 013 

\bibitem{bertschinger} 
E. Bertschinger \& P. Zukin, arXiv:0801.2431 

\bibitem{DvaliTurn03} 
G. Dvali \& M.S. Turner 2003, arXiv:astro-ph/0301510

\bibitem{freese03} 
P. Gondolo \& K. Freese 2003, Phys. Rev. D 68, 063509 

\bibitem{dgp}
G. Dvali, G. Gabadadze, M. Porrati 2000, Phys.\ Lett.\ B 485, 208

\bibitem{ddg}
C. Deffayet, G. Dvali, G. Gabadadze 2002, Phys. Rev. D 65, 044023 

\bibitem{scherrer06} 
R.J. Scherrer 2006, Phys. Rev. D 73, 043502 

\bibitem{songhs} 
Y-S. Song, W. Hu, I. Sawicki 2007, Phys. Rev. D 75, 044004 

\end{thebibliography}
\end{document}